# Search for the cyclic activity on red dwarfs from photometric surveys

*Bondar' N.I.[1], Gorbunov M.A.[2] and Shlyapnikov A.A.[3]*

[1]*Crimean Astrophysical Observatory RAS, Nauchny, Crimea, Russia; otbn@mail.ru*
[2]*Crimean Astrophysical Observatory RAS, Nauchny, Crimea, Russia; mag@craocrimea.ru*
[3]*Crimean Astrophysical Observatory RAS, Nauchny, Crimea, Russia; aas@craocrimea.ru*

**Abstract.** Modern databases and rows of observations allow to make progress in study of magnetic activity of solar-type stars and cold dwarfs. Digitizing of wide-filed plates in the large archives give historical data expanding the investigated time intervals up to century. Photometric sources for a seek and study of cycles produced by starspots are considered and results for several red dwarfs are presented.

## 1. Introduction

The study of the magnetic activity of stars requires a long series of observations of activity indicators in different layers of the stellar atmospheres. This follows from the experience of studying solar cycles, the nature and features of the development of one type of them, the Schwabe cycle, are studied for more than 3 centuries. Non-stationary phenomena on low-mass main sequence stars and the Sun have a phenomenological similarity and manifest themselves in the form of chromospheric flares, photospheric spots, and changes in coronal activity. Among G-M dwarfs, there are distinguish stars with a predominant flare activity, as on the M-star UV Cet, and with a small amplitude rotational modulation, typical to the K dwarf BY Dra (Gershberg, 2005).

The first evidence of cyclic activity in F-K dwarfs was obtained from decade observations of the intensity of chromospheric lines H and K of CA II (Wilson, 1978). Continuing of these research for 111 stars in the framework of the HK project during 25 years Baliunas et al. (1995) showed that stellar cycles can be longer or shorter than the 11-year solar cycle of activity. At present, cycles of the chromospheric activity on 29 stars from the Maunt Wilson (MW) project has been studied for an interval of 36 years (Oláh et al., 2016).

The cycles associated with changes in photospheric brightness produced by starspots were found from photographic collections. A duration of a cycle on some stars exceeds some decades. However, photographic series have low accuracy and data filling; it is necessary to use several collections for confidence in the result. Currently, many observatories scanned archival records and provided access to files by means of virtual observatories. Information about photographic archives is collected in the photographic data center in Sofia, a description of this database is given in (Tsvetkov et al., 2001). Since the beginning of the 1990s, programs for monitoring the brightness of active dwarfs using automated telescopes are conducted (Strassmeier, 2005), and since 2000, wide-field CCD patrols, such as ASAS (Pojmanski, 1997). Large data volumes are contained in the archives of space projects Hipparcos, Kepler, Catalina, GAIA.

In this paper, for some active dwarfs, the results of a seek and study of cycles on long-term photometric series, which have formed from photometric catalogues and surveys are considered.

## 2. Photometric sources for study of stellar activity

Methods of photometry allow us to study the surface manifestations of activity caused by the development of spots. Changes in the intensity of the process of formation of spots can be both irregular and cyclical. The main cycle (cycle with a highest amplitude) can last several years or even decades. Photographic collections allow us to study the behavior of brightness of stars over long time intervals, they are a resource for searching for the main cycles of a large number of stars. The use of different archives leads to an increasing the reliability of the obtained light curves, but their accuracy of $0^m.05$–$0^m.1$ makes it possible to detect cycles only in stars with significant spottedness of the photosphere. Difficulties arise in the reduction of the measured data. For individual stars Phillips & Hartmann (1978) have made special photoelectric observations of reference stars in filter B. Bondar', 1995 investigated the brightness of 40 active dwarfs on plates in the archives of the Sternberg State Institute (MSU), Odessa and Sonneberg Observatories on time spans of several decades, and 12 stars of them with the supposed cycles were selected for further photoelectric observations.

Digitizing of wide-field plates makes a work with information more productive, but data accuracy remains photographic. In the study of cycles, photographic measurements are used in complete with data from modern photometry to construct complete light curves.

The ASAS photometric catalog data was obtained for stars with declination $\delta < 28°$. In the V-filter stars of $8^m - 12^m$ are available for research starting from 2000. Each value is given with an error and inaccurate data can be eliminated during the initial processing. This catalog has been used to search for cycles for a large number of cold dwarfs, but the research interval is about of 9 years so far (Suarez Mascareno et al., 2016). We search for long-term cycles from compilative light curves based on photographic measurements, ASAS, SuperWASP, KWS databases and obtained by the Hipparcos satellite, as well as own and published photometric results.

This approach made it possible to find the longest spots cycles, up to 80 years, on V833 Tau (dK5) (Bondar', 2015) and V647 Her (dM3.5) (Bondar', 2018a), and long-term cycles on M dwarfs YZ CMi (Bondar', Katsova, 2018b), DT Vir, V577 Mon (Bondar', 2018a). In Fig.1. the summary light curve of the V833 Tau shows the cyclic changes in annual average values with a characteristic time of 78.25 years.

But it is obviously that a complete character of variations in light curve does not described by a simple cycle. In paper (Bondar', 2015) we found also 19-year cycle, it is stable, but its maxima (amplitudes) and form vary from cycle to cycle. It is not excluded that the cycle of 78 year is similar to the 80-year solar cycle, and shown grand maxima of activity, and the 19-year cycle is the main one.

The similarity of the development of cyclic activity on this star with photospheric cycles on the Sun is noted by Oláh et al. (2009). The multiple cycles, which is feature of solar activity, are found also on this K-star. In addition to the long cycle, short cycles of 6.24 and 2.45 years were found, associated with the evolution of spots and their distribution. We also confirmed the existence of these cycles (Bondar', 2015).

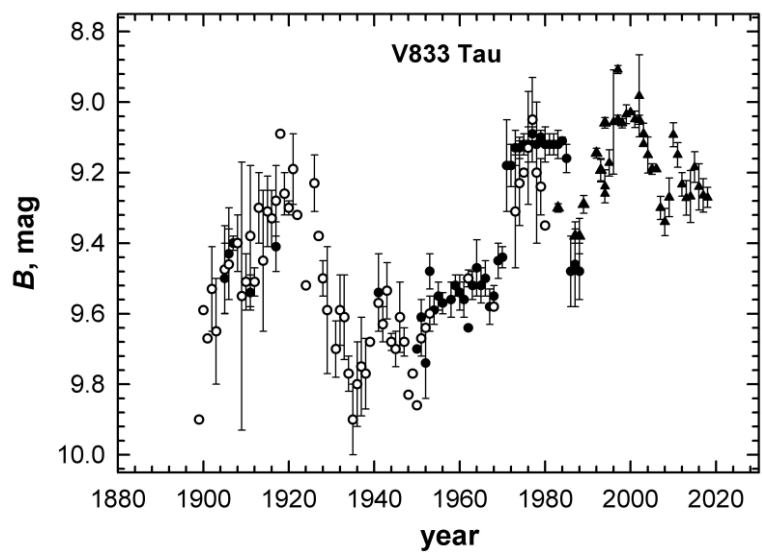

*Figure 1.* Light curve of V833 Tau and long-term cycle. Left: compilative yearly mean magnitudes obtained from the Harvard photographic archive by Hartmann et al., 1981 – open circles, by Bondar', 1995 – black circles, black triangles mark photometric data from Oláh, Pettersen, 1991, Strassmeier et al., 1997, Oláh et al., 2001, Bondar', 2015, ASAS and KWS catalogues.

The catalogues ASAS and SuperWASP contain data obtained with temporal resolution which allow to study the distribution of spots and to seek cycles associated with changes of active longitude (flip-flop cycles). Figure 2 shows changes in location of active region on surface of YZ CMi since 2003 to 2009 (Bondar', 2018b).

Long-term changes of photometric period are indicators of migration of starspots, which may be cyclic in analogy to the known on the Sun. Chugainov (1974) found such changes of photometric period from many years observations of BY Dra. Working with ASAS and SuperWASP databases variations of photometric period on long-term scale may be study for many solar – type stars and active red dwarfs. High temporal resolution of SuperWASP data allow to evaluate the differential rotation of the star and find fast-rotating stars with an axial rotation of less than a day. Using this catalog we found that V647 Her is a fast-rotating cold dwarf with a period of 1.09 days, and that the rotation periods of V374 Peg (0.44 d) and DT Vir (2.88d) have not changed during many years (Bondar', 2018).

The KWS catalog presents results since 2000 and is a source of data in VI-filters, which is very important for finding spots, determining their parameters and building models for their distribution.

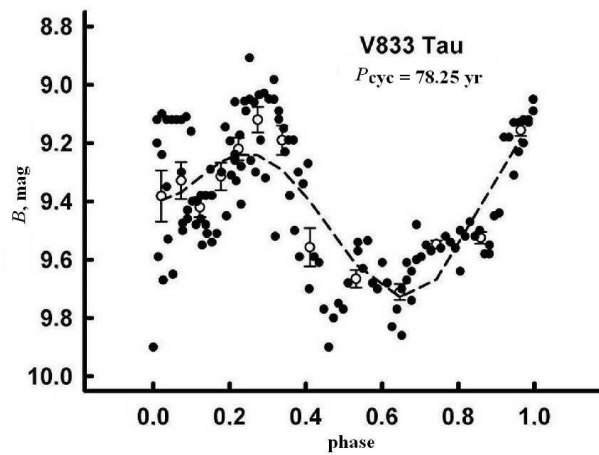

*Figure 2.* Cycle in changes of yearly mean brightness (black circles) is fitted by a high degree polynomial drawn by dashed through average values in bins of 0.1 phase Pcyc = of 78.25 yr (open circles).

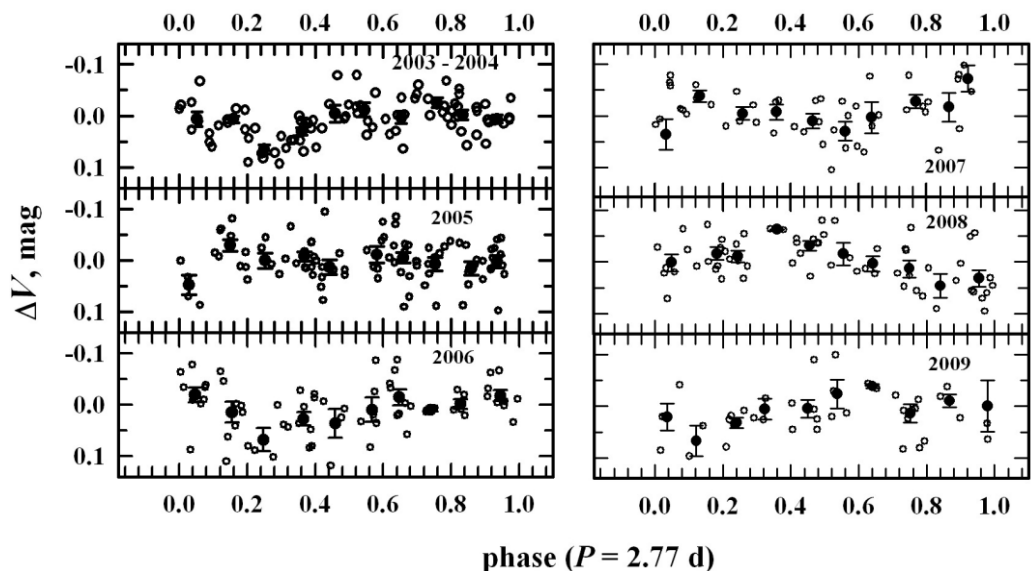

*Figure 3.* Longitudinal changes in location of active region on photosphere of M dwarf YZ CMi in 2003–2009. Open circles – ASAS data, black circles – the averaged in bins of 0.1 phase Prot = 2.77 day.

## 3. Search for cycles of photometric activity for selected red dwarfs

One of the sources to preparation the program list for study of cyclic activity on red dwarfs is the GTSh10 catalog (http://craocrimea.ru/~aas/CATALOGUEs/G+2010/eCat/G+ 2010.html). It contains information for a large number red dwarfs which the brightness changes produced by photospheric spots. We prepared a list of these stars in the required format for obtaining their V magnitudes from the CSDR-2 database (Drake et al., 2009) and selected 129 objects with the photometric series of data more than 8 years. Examples of long-term changes in light curves possible associated with the manifestation of cyclic activity are shown in Figures 4.

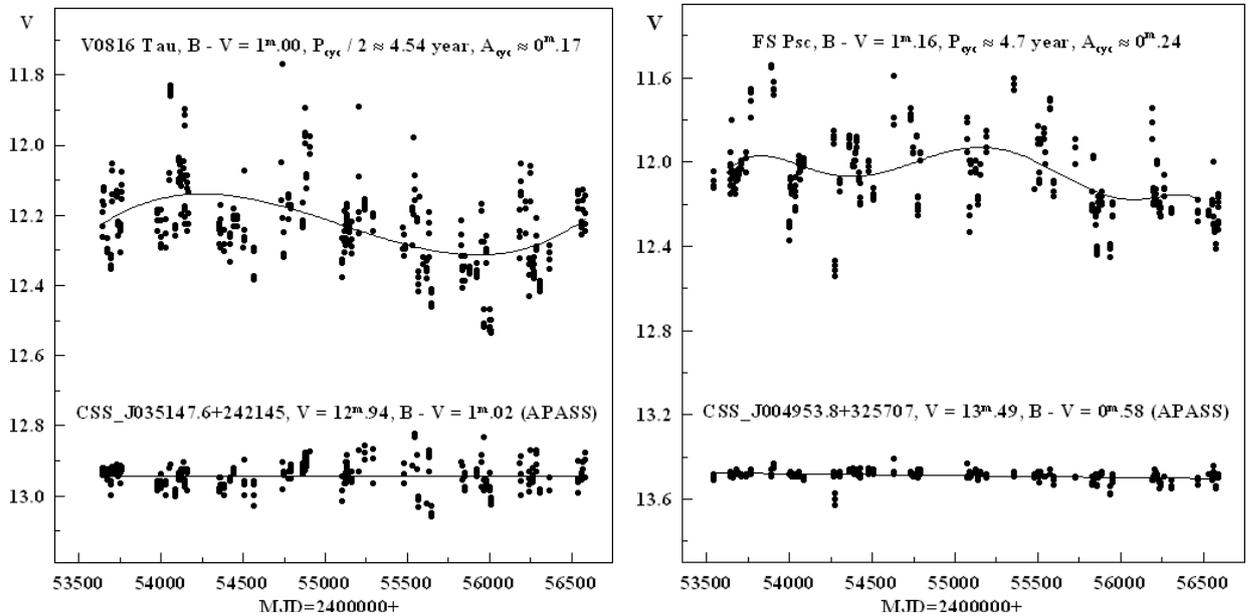

*Figure 4*. Long-term variations in brightness of V816 Tau (left) and FS Psc (right) probable related to a cyclic starspots activity.

The variability of V0816 Tau star was discovered and classified as BY Dra-type by Meys et al. (1982). In this publication the star is designated as HZ 3163. We used the comparison star CSS_J035147.6 + 242145 from CSDR-2, separated from V0816 Tau at a distance of ~ 2′ and having a similar color index to exclude the trend not related to the change in the brightness of V0816 Tau. As can be seen from the figure 4, an object may have half the cyclic brightness change with a duration of approximately 4.54 years and the peak-to-peak amplitude of $0^m.17$ in change of the averaged brightness.

The star FS Psc was discovered and classified as BY Dra type by Bernhard & Lloyd (2008) at program of optical identification of X-ray sources from the ROSAT All-Sky Faint Source Catalogue (1RXS). We present light curve of the star and the comparison star CSS_J004953.8 + 325707 from CSDR-2, separated from FS Psc at a distance of ~ 1′.3. According Figure 4 the whole cycle of variations in V-values is approximately of 4.7 years and an amplitude of variability exceeds of $0^m.24$ magnitude. There is also some trend in dimming over time.

## 4. Conclusions

The present observations of indicators of activity on solar-type stars and cold dwarfs cover time span of some decades only for relative small number of objects. Cycles comparable with the solar one are found more certainly. The long-term cycles with duration more than 20 year should be confirmed by further observations. Digitizing of wide-field plates in many photographic archives give historical data for supplementing of photometric series for many stars. Such approach allow us to reveal objects with the suspected long cycles. Modern ground and space surveys and created databases allow to search for cycles of magnetic activity for a large number objects with different physical parameters and study stellar activity on different evolutionary stages.

**Acknowledgments.** The authors are grateful to Z. Taloverova for the preparation of the article to publication. This work reported at the conference "Physics of Magnetic Stars" (1–5 October, 2018, Special Astrophysical Observatory, Russia) and we thankful for its organization.

The second author is grateful for partial support of this work by the Russian Foundation for Basic Research. The reported study was funded by RFBR according to the research project 18-32-00775.